\documentstyle[12pt]{article}
\title{Proper co-ordinates of non-inertial observers and rotation}
\author{Hrvoje Nikoli\'c  \\
Theoretical Physics Division, Rudjer Bo\v{s}kovi\'{c} Institute, \\
P.O.B. 180, HR-10002 Zagreb, Croatia \\
{\normalsize hrvoje@thphys.irb.hr} \\
\makebox[1in]{} \\
}
\date{\today}
\begin{document}
\maketitle
\begin{abstract}
By proper co-ordinates of non-inertial observers 
(shortly - proper non-inertial co-ordinates)
we understand the proper co-ordinates of an
arbitrarily moving local observer. 
After a brief review of the theory of proper non-inertial co-ordinates, 
we apply these co-ordinates to discuss the relativistic 
effects seen by observers at different positions on a 
rotating ring. 
Although there is no relative motion among  
observers at different positions, they belong to different 
proper non-inertial frames. 
The relativistic 
length seen by an observer depends only on his instantaneous 
velocity, not on his acceleration or rotation. 
For any observer
the velocity of light is isotropic and 
equal to $c$, provided that it is measured by propagating a light
beam in a small neighbourhood of the observer. 
\end{abstract}

\section{Proper non-inertial co-ordinates}

In physics, all dynamical equations of motion are certain 
differential equations that describe certain quantities as 
functions of space-time points. Space-time points 
are parametrized by their co-ordinates. It is convenient to 
write the equations of motion (as well as other related 
equations) in a form which is manifestly covariant with 
respect to general co-ordinate transformations. When one 
solves the equations, one must use some specific co-ordinates. 
The covariance provides that one can use any co-ordinates 
he wants, because later he can easily transform the 
results to any other co-ordinates. Therefore, it is convenient 
to choose co-ordinates that simplify the technicalities of the
physical problem considered.  

The general covariance is often interpreted as a statement 
that ``physics does not depend on the co-ordinates chosen".
However, this is not so. The choice of   
co-ordinates is more than a matter of convenience. 
The main purpose of
theoretical physics is to predict what will be {\em observed} under
given circumstances. 
The main lesson we have learned from Lorentz 
co-ordinates is the fact that what an observer observes
(time intervals, space intervals, components of a tensor, ...)
depends on how the observer moves. Lorentz co-ordinates 
are proper co-ordinates of an observer that moves inertially in 
flat space-time. Proper non-inertial co-ordinates are the 
generalization of Lorentz co-ordinates to arbitrary motion 
in arbitrary space-time. If one is interested in how 
a physical system  
looks like to a specific observer, one must transform the results to the
corresponding proper non-inertial co-ordinates. 

Proper non-inertial co-ordinates are determined by the (time-like) trajectory of 
the observer, by the rotation of the observer with respect to a local 
inertial observer and by the properties of space-time 
itself. The general geometrical construction of proper non-inertial co-ordinates is 
well established \cite{mtw}. Here I present the most important properties 
of proper non-inertial co-ordinates:
\begin{enumerate}
\item Proper non-inertial co-ordinates are chosen such that the position of the 
observer is given by $x^{\mu}=(t,0,0,0)$, where $t$ is the time measured 
by a clock co-moving with the observer.
\item The metric expressed in proper non-inertial co-ordinates has the property
 \begin{equation}
 g_{\mu\nu}(t,0,0,0)=\eta_{\mu\nu} \; .
 \end{equation}
\item The connections $\Gamma^{\alpha}_{\beta\gamma}$ vanish at 
 $(t,0,0,0)$ if and only if the trajectory is a geodesic and 
 there is no rotation.
\end{enumerate}

The general geometrical construction of proper non-inertial co-ordinates is 
not very useful for practical calculations. However, in 
flat space-time, proper non-inertial co-ordinates can be constructed in an alternative way,  
more useful for practical calculations \cite{nels}. 
Here I present an elegant form of this construction 
introduced in \cite{nikolic1}.
 
Let $S$ be a Lorentz frame 
and let $S'$ be the proper non-inertial frame of the observer whose 3-velocity is
$u^i(t')\equiv \mbox{\bf{u}}(t')$, as seen by an observer in $S$.
The co-ordinate transformation between these two frames 
can be obtained by summing the infinitesimal Lorentz transformations.
Let
\begin{equation}\label{el1}
x^{\mu}=f^{\mu}(t',\mbox{\bf{x}}';\mbox{\bf{u}})
\end{equation}
denotes the ordinary Lorentz transformation, i.e. the transformation
between two Lorentz frames specified by the constant relative velocity
$\mbox{\bf{u}}$.
The quantities 
\begin{equation}\label{partial}
f^{\mu}_{\; \nu}=\left( \frac{\partial f^{\mu}}{\partial x'^{\nu}}
 \right)_{\mbox{\bf{u}}=\mbox{\rm{const}}} 
\end{equation}
have explicit values
\begin{eqnarray}\label{fmini}
& f^{0}_{\; 0}=\gamma \; , \;\;\;\;\; f^{0}_{\; j}=-\gamma u_j \; ,
\;\;\;\;\;
  f^{i}_{\; 0}=\gamma u^i \; , & \nonumber \\
& f^{i}_{\; j}=\delta^{i}_{\; j}+
 \displaystyle\frac{1-\gamma}{\mbox{\bf{u}}^2} u^i u_j \; , &
\end{eqnarray}
where $u^j =-u_j$, $\mbox{\bf{u}}^2 =u^i u^i$,
$\gamma=1/\sqrt{1-\mbox{\bf{u}}^2}$. 
The differential of (\ref{el1}) is
\begin{equation}\label{pra13}
dx^{\mu}=f^{\mu}_{\; \nu}(t',\mbox{\bf{x}}';\mbox{\bf{u}}) dx'^{\nu}.
\end{equation}
The transition to a noninertial frame introduces 
a time-dependent velocity: $\mbox{\bf{u}}\rightarrow\mbox{\bf{u}}(t')$.
The transformation between $S$ and $S'$ is given by the integration of 
(\ref{pra13}) in the following way:
\begin{equation}\label{el3}
x^{\mu}=\int_{0}^{t'}f^{\mu}_{\; 0} (t',0;\mbox{\bf{u}}(t')) dt' +
\int_{C}
f^{\mu}_{\; i} (t',\mbox{\bf{x}}';\mbox{\bf{u}}(t')) dx'^{i} \; .
\end{equation}
In (\ref{el3}), $C$ is an arbitrary curve with constant          
$t'$, starting from $0$    
and ending at $x'^{i}$.

In general, $S'$ can also rotate i.e. change the orientation
of the co-ordinate axes with respect to an inertial frame.
The rotation can be described by
the rotation matrix $A_{ji}(t')=-A_{j}^{\; i}(t')$.
It
satisfies the differential equation
\begin{equation}\label{er4}
 \frac{d A_{ij}}{dt}=-A_{i}^{\; k}\omega_{kj} \; ,
\end{equation}
where $\omega_{ik}=\varepsilon_{ikl}\omega^{l}$, $\varepsilon_{123}=1$  
and $\omega^i(t')$ is the angular velocity as seen by an observer in $S$. 
In this more general case the transformation is also given by 
(\ref{el3}), but now $C$ is an arbitrary curve with constant
$t'$, starting from $0$
and ending at $-A_{j}^{\; i}(t')x'^{j}$.
Note that the proper non-inertial co-ordinates $x'^{\mu}$ 
are constructed such that 
the space origins of $S$ and $S'$ coincide at $t=t'=0$. 
 
The metric tensor in $S'$ is given by 
\begin{equation}
g'_{\mu\nu}=\frac{\partial x^{\alpha}}{\partial x'^{\mu}}
\frac{\partial x^{\beta}}{\partial x'^{\nu}}
g_{\alpha\beta} \; ,
\end{equation}
where $g_{\alpha\beta}=\eta_{\alpha\beta}=
{\rm diag}(1,-1,-1,-1)$ is the Minkowski 
metric in $S$.
Using (\ref{el3}), it is straightforward 
(but slightly tedious) to obtain  
\begin{eqnarray}\label{metric}
 & g'_{ij}=-\delta_{ij} \; , \;\;\;\;\;
   g'_{0j}=-(\mbox{\boldmath $\omega$}'\times\mbox{\bf{x}}')_j \; , &
\nonumber \\
 & g'_{00}=c^2 \left(
1+\displaystyle\frac{\mbox{\bf{a}}'\cdot\mbox{\bf{x}}'}{c^2}
    \right)^2 -(\mbox{\boldmath $\omega$}'\times\mbox{\bf{x}}')^2 \; , &
\end{eqnarray}
where
\begin{equation}
 \omega'^i =\gamma (\omega^i -\Omega^i) \; , \;\;\;\;\;
 a'^i =\gamma^2 \left[a^i +\frac{1}{\mbox{\bf{u}}^2}(\gamma
  -1)(\mbox{\bf{u}}\cdot\mbox{\bf{a}})u^i\right] \; ,
\end{equation}
$\Omega^{i}$ is the time-dependent Thomas precession frequency
\begin{equation}
\Omega_{i}=\frac{1}{2\mbox{\bf{u}}^2}(\gamma -1)\varepsilon_{ikj}
 (u^k a^j -u^j a^k) \; ,
\end{equation}
and $a^i=du^i/dt$ is the time-dependent acceleration.
 
From Property 2 we see that the space co-ordinates $x^i$ are a 
generalization of Cartesian co-ordinates. However, this does not 
imply that an observer is not allowed to use polar co-ordinates, 
for example. The most general co-ordinate transformations that 
correspond merely to a redefinition of the co-ordinates of
the same physical observer are the so-called restricted internal transformations 
\cite{nikolic1}, i.e. the transformations of the form 
\begin{equation}\label{veryweak}
t'=f^0 (t) \; , \;\;\;\; x'^i=f^i (x^1,x^2,x^3) \; , 
\end{equation}
where $t,x^i$ are proper non-inertial co-ordinates. The quantities $g_{00}dt^2$ and 
\begin{equation}\label{dl}
dl^2=-g_{ij}dx^{i}dx^{j} \; 
\end{equation} 
do not change under restricted internal transformations. In order 
to describe physical effects as seen by a local observer, one must 
use proper non-inertial co-ordinates modulo restricted internal transformations. 
For simplicity, in the rest of the paper I use proper non-inertial co-ordinates. 

Two observers with different trajectories have different proper non-inertial frames. 
In particular, this implies that {\em even if there is no relative motion 
between two observers, they belong to different frames if they do not 
have the same position}. This fact was not realized 
in many previous papers, which led to many misinterpretations. 
At first sight, this fact contradicts the well-known fact that 
two inertial observers in flat space-time may be regarded as 
belonging to the same 
Lorentz frame if there is no relative motion between them. However, this is 
because their proper non-inertial frames (with the space origins at their positions) 
are related by a translation of the space origin, which is a 
restricted internal transformation. In general, for practical purposes, 
two observers can be regarded  
as belonging to the same proper non-inertial frame if there is no relative motion between
them and the other observer is close 
enough to the first one, in the 
sense that the metric expressed in proper non-inertial co-ordinates 
of the first observer does not depart
significantly from $\eta_{\mu\nu}$ at the position of the second 
one. It is an exclusive
property of Minkowski co-ordinates, among other
proper non-inertial co-ordinates in flat or curved space-time, that the
metric is equal to $\eta_{\mu\nu}$ {\em everywhere}.
This implies that
two observers at different positions but
with zero relative velocity may be regarded as
belonging to the same co-ordinate frame only if they
move inertially in flat space-time.  
  
\section{Application to rotation}

For motivation, let us first review the problems \cite{nikolic1}
of the standard resolution \cite{gron1,gron2} of the Ehrenfest paradox.  
We study a rotating ring in a rigid non-rotating
circular gutter with radius $r$.
One introduces the co-ordinates
of the rotating frame $S'$
\begin{equation}\label{eq1}
 \varphi'=\varphi -\omega t \; , \;\;\;\; r'=r \; , \;\;\;\; z'=z
  \; , \;\;\;\; t'=t \; ,
\end{equation}
where $\varphi$, $r$, $z$, $t$ are cylindrical co-ordinates of the
inertial frame $S$ and $\omega$ is the angular velocity. The metric in $S'$
is given by
\begin{equation}\label{eq2}
 ds^2=(c^2-\omega^2 r'^2)dt'^2 -2\omega r'^2 \, d\varphi' dt' -dr'^2
 -r'^2 \, d\varphi'^2 -dz'^2 \; .
\end{equation}
It is generally accepted that the space line element should be
calculated by the formula \cite{land}
\begin{equation}\label{eq3}
 dl'^2=\gamma'_{ij}dx'^i dx'^j \; , \;\;\; i,j=1,2,3 \; ,
\end{equation}
where
\begin{equation}\label{eq4}
 \gamma'_{ij}=\frac{g'_{0i}g'_{0j}}{g'_{00}}-g'_{ij} \; .
\end{equation}
This leads to the circumference of the ring 
\begin{equation}\label{eq5}
 L'=\int_{0}^{2\pi} \frac{r' d\varphi'}{\sqrt{1-\omega^2 r'^2/c^2}}
   =\frac{2\pi r'}{\sqrt{1-\omega^2 r'^2/c^2}}   
   \equiv\gamma(r')2\pi r' \; .
\end{equation}
The circumference of the same ring as seen from $S$ is $L=2\pi r=2\pi r'$.
Since the ring is constrained  to have the same radius $r$ as the same 
ring when it does not rotate, $L$ is not changed by the rotation, but
the proper circumference $L'$ is larger than the proper circumference of
the non-rotating ring. This implies that there are tensile stresses in the
rotating ring. The problem is that it is assumed here that (\ref{eq1}) 
defines the proper frame of the whole ring. This
implies that an observer on the ring sees that the 
circumference is $L'=\gamma L$. The circumference of the gutter
seen by him 
cannot be different from the circumference of the ring
seen by him, so the 
observer on the ring sees that the circumference of the
relatively moving gutter is {\em larger} than the proper
circumference of the gutter, whereas we expect,
using our experience with the usual Lorentz 
contraction, that he should
see that it is smaller. Of course, it depends on how 
the circumference is measured. Here we have in mind an 
experimental procedure that in principle can also be used 
to measure the usual Lorentz contraction, based on  
photographing with a very short exposition, such that 
the change of the photographed object position during the 
exposition can be neglected. The size of the object's picture 
on the photography corresponds to the measured size. Obviously, 
with such a measuring procedure, for any observer the 
apparent circumference of the {\em whole} ring must be 
equal to the apparent 
circumference of the {\em whole} gutter. 
This is a simple consequence of the 
fact that, at any instance of time, any part of the ring is 
somewhere inside the gutter and any part of the gutter has a 
part of the ring near it.
(Note also that this is not so 
for a well known ``paradox" of a car in a garrage where different 
observers may disagree on whether a fast car can fit into an {\em open} 
garrage at rest. This is because, for each part of the car, there are 
times for which that part is outside the garrage as well as times 
for which it is inside the garrage.)  

The problem discussed above
resolves when one realizes that (\ref{eq1}) does not define 
the proper frame of the ring. Each point on the ring belongs to 
a different proper non-inertial frame. The co-ordinates (\ref{eq1}) are actually 
proper non-inertial co-ordinates (modulo a restricted internal transformation) of an 
observer that rotates in the centre of the ring. 
However, this raises another 
problem. If (\ref{eq3}) is the correct definition
of the space line element, then the observer that rotates in the centre 
should see that the circumference
of the gutter is larger 
than the proper circumference of the gutter by a factor $\gamma(r')$.
However, $\omega r'/c$ can be 
arbitrarily large, so $\gamma(r')$ can be not only 
arbitrarily large, but also even imaginary. On the other hand, 
we know from
everyday experience that the apparent velocity $\omega r'$ of stars,
due to our rotation,  
can exceed the velocity of light, but we see neither a contraction 
nor an elongation of the stars observed. This implies that the 
definition of the space line element (\ref{eq3}) is not always 
appropriate. In \cite{land}, (\ref{eq3}) 
is derived by defining the space line element 
through a measuring procedure that lasts a finite time, so, in general, 
this 
formula is not appropriate for a definition of the {\em instantaneous} 
length.
Of course, if rotation is stationary, then it is not necessary 
to use a definition appropriate for the instantaneous length.
However, we want to have a definition that is appropriate for {\em any} 
case and to consider a stationary rotation only as a special case 
of arbitrary motion.  
In flat space-time, as shown in \cite{nikolic1},  
if physics is described by proper non-inertial co-ordinates modulo restricted internal
transformations, a more appropriate definition of the space line element 
is (\ref{dl}).  
 
Using the co-ordinate transformation (\ref{el3}), 
one can study the relativistic contraction in the same way 
as in the conventional approach with Lorentz frames. One assumes that 
two ends of a body are seen to have the same time co-ordinate. From 
(\ref{el3}) and (\ref{fmini}) one can easily see that the co-ordinate 
transformation is linear in $x'^i$. As demonstrated in more detail in 
\cite{nikolic1,nikolic3}, it implies that {\em an arbitrarily accelerated and
rotating
observer sees equal lengths of other differently
moving objects as an inertial observer whose
instantaneous position and velocity are equal to that of the
arbitrarily accelerated and rotating observer.}
  
Using (\ref{el3}), one can also study the rate of clocks as seen by 
various observers. In particular, one can study the twin paradox for 
various motions of the observers \cite{nikolic4}. 
However, it is more interesting to 
study not only the time shift after the two differently moving 
observers eventually meet, but also the continuous changes of the 
time shifts during the motion. As can be seen from (\ref{el3}), 
it is the time dependence (not the space dependence) of the 
co-ordinate transformation that significantly differs from 
the ordinary Lorentz transformations. One cannot invert (\ref{el3}) 
simply by putting $u^i \rightarrow -u^i$. Therefore, inertial 
and non-inertial observers see quite different continuous changes of the 
time shift. 

Following \cite{nikolic1}, let us discuss in more detail the case of 
uniform circular motion.
Here we only present the results, while the technical details 
are delegated to the Appendix.
Assume that there are three clocks. One is at rest in $S$, 
so it moves inertially. The other two are moving around a circle with the 
radius $R$ and have the velocity $\omega R$ in the 
counter-clockwise direction, as seen by the observer in 
$S$. The relative angular distance between these two non-inertial clocks 
is $\Delta\varphi_0$, as seen in $S$. The inertial observer will see that 
the two non-inertial clocks lapse equally fast, so I choose that he sees 
that they show the same time. He will see the clock rate $t=\gamma t'$, 
where $\gamma=1/\sqrt{1-\omega^2 R^2/c^2}$. The observer co-moving with 
one of the non-inertial 
clocks will not see that the other non-inertial clock shows the 
same time as his clock. 
He will see the constant time shift given by the equation
\begin{equation}\label{eqgron1}
\gamma\omega (t''-t')=\beta^2\sin (\gamma\omega
(t''-t')+\Delta\varphi_0) \; ,
\end{equation}
where $\beta^2 \equiv \omega^2 R^2 /c^2$ and $t''$ is the time of the other 
non-inertial clock.
Finally, let us see how the inertial clock looks like from the point 
of view of the observer co-moving with one of the non-inertial clocks. 
Let the position of the inertial clock be given by its co-ordinates 
$(x,y)$. We choose the origin of $S$ such that, at $t=t'=0$, the 
space origins of $S$ and $S'$ coincide  
and the velocity of the non-inertial 
observer is in the $y$-direction, as seen in $S$. The rate of 
clocks as seen by the non-inertial observer is given by
\begin{equation}\label{txy}
t=\gamma t' + 
 \frac{\omega R}{c^2}[y \cos \gamma \omega t'
 -(x+R) \sin \gamma \omega t'] \; .
\end{equation}
The oscillatory functions in (\ref{txy}) vanish when
they are averaged over time. 
This means that the observer in $S'$ agrees
with the observer in $S$ that the clock in $S'$ is slower, but only in
a time-averaged sense. For example, when these two clocks 
are very close to each other, 
then, by expanding (\ref{txy}) for small $t'$, one finds $t=t'/\gamma$, 
which is the result that one would obtain if the velocity of the
non-inertial  
observer were constant.  
If the clock in $S$ is in the centre, which
corresponds
to $x=-R$, $y=0$, then (\ref{txy}) gives $t=\gamma t'$, so in this case
there is no oscillatory behaviour.     

Finally, let us shortly discuss the implications on the velocity 
of light.
If (\ref{eq1}) is interpreted as a proper frame of all observers on a 
rotating platform, then one can conclude that the observer on the 
rotating platform will see that the velocity of light depends on 
whether light is propagating in the clockwise or in the counter-clockwise 
direction (see, for example, \cite{kla}). This is related to the 
fact that the metric (\ref{eq2}) is not time orthogonal.  
However, now we know that each observer
belongs to a different proper non-inertial frame, and from Property 2 we see
that
in the {\em vicinity} of any observer the metric is equal to the
Minkowski metric $\eta_{\mu\nu}$. This implies that {\em
for any local observer the velocity of light is isotropic and is
equal to $c$, provided that it is measured by propagating a light
beam in a {\bf small} neighbourhood of the observer.} 
In particular, this leads to a 
slightly different interpretation of the Sagnac effect \cite{nikolic1}, 
in which the velocity of light as seen by the observer on the rotating 
platform depends on the instantaneous relative position of the light 
beam with respect to the observer. The details are presented in the 
Appendix.  

\section*{Acknowledgments}

The author is grateful to G. Rizzi for his editorial critical 
remarks that significantly improved this paper.
This work was supported by the Ministry of Science and Technology of the
Republic of Croatia under the contract No. 0098002.

\section*{Appendix}

In this Appendix, we derive Eqs. (\ref{eqgron1}) and (\ref{txy}) 
and discuss how the velocity of light apppears to an observer 
on the rim of a rotating platform. 
In these derivations, we follow \cite{nikolic1}.

Let the gutter be placed at the $z=0$ plane. We put the
space origin of $S$ at a fixed point on the gutter, such that
the $y$-axis is tangential to the gutter and the $x$-axis is perpendicular
to the gutter at $\mbox{\bf{x}}=0$. In the rest of the analysis
the $z$-co-ordinate can be  suppressed. A part of the ring in the gutter
can be
considered as a short rod initially 
placed at $\mbox{\bf{x}}=0$ and
uniformly moving along the gutter in the counterclockwise direction.
The gutter causes a torque that provides that the rod is
always directed tangentially to the gutter. Therefore,
$\omega=u/R$, where $u=\sqrt{\mbox{\bf{u}}^2}$ is the time independent 
speed of rod.
Now, $\gamma=1/\sqrt{1-\omega^2 R^2/c^2}$ is also time independent.
Since a clock in $S'$ is at $\mbox{\bf{x}}'=0$, the clock rate between
a clock in $S$ and a clock in $S'$ is given by $t=\gamma t'$, as seen
by an observer in $S$. We assume that, initially, the axes $x'$, $y'$ are
parallel to the axes $x$, $y$, respectively. Therefore the velocity
\begin{equation}
\mbox{\bf{u}}(t')=\omega R (-\sin \gamma\omega t', \cos \gamma\omega t')
\end{equation}
is always in the $y'$-direction and the solution of (\ref{er4}) is
\begin{equation}
A_{ij}(t')=\left(
 \begin{array}{cc}
  \cos \gamma\omega t' & \sin \gamma\omega t' \\
  -\sin \gamma\omega t' & \cos \gamma\omega t'
 \end{array} \right) \; .
\end{equation}
The transformation (\ref{el3}) reduces to
\begin{equation}\label{er1'}
 \left( \begin{array}{c} x \\
                         y
        \end{array} \right)=
 \left( \begin{array}{cc}
   \cos \gamma\omega t' & -\gamma\sin \gamma\omega t' \\
   \sin \gamma\omega t' & \gamma\cos \gamma\omega t'
  \end{array} \right)
 \left( \begin{array}{c} x' \\
                         y'
        \end{array} \right)
 +R \left( \begin{array}{c} \cos \gamma\omega t' -1 \\
                            \sin \gamma\omega t'
           \end{array} \right) \; ,
\end{equation}
\begin{equation}\label{er2'}
t=\gamma t' +\frac{\gamma}{c^2}\omega R y' \; .
\end{equation}
In particular, at $t'=0$ these transformations become
\begin{equation}\label{t=0}
 x=x' \; , \;\;\;\; y=\gamma y' \; , \;\;\;\; t=\frac{\gamma u}{c^2}y' \; ,
\end{equation}
which coincide with the ordinary Lorentz boost at $t'=0$ for the velocity
in the $y$-direction.

Let us now study how other parts of the ring
appear to the observer on the ring. 
Since the rotation is uniform
the result cannot depend on $t'$, so without losing 
on generality, we evaluate this at
$t'=0$.
We introduce polar
co-ordinates $r$, $\varphi$, defined by
\begin{equation}
y=r \sin \varphi \; , \;\;\;\; R+x=r \cos \varphi \; ,
\end{equation}
which are new space co-ordinates for $S$, with the origin in the center
of the circular gutter. The angle $\varphi$ is a good label of the 
position
of any part of the ring even in $S'$. (To visualize this, one can draw
angular marks on the gutter. The number of marks separating two points
on the gutter or on the ring is a measure of the ``angular distance" in
any frame.)
Let $S''$ be the frame of another
part of the ring. The position of that part of the ring is
$x''=y''=0$. The relative position of the space origin of $S''$ with 
respect
to that of $S'$ is given by the constant relative angle $\Delta\varphi_0$,
as seen by an observer in $S$. In analogy
with (\ref{er1'})-(\ref{er2'}), we find that $S''$ is determined by
\begin{eqnarray}\label{er1''}
 \left( \begin{array}{c} x \\
                         y
        \end{array} \right) & = &
 \left( \begin{array}{cc}
   \cos (\gamma\omega t''+\Delta\varphi_0) & -\gamma\sin (\gamma\omega
          t''+\Delta\varphi_0) \\
   \sin (\gamma\omega t''+\Delta\varphi_0) & \gamma\cos (\gamma\omega
          t''+\Delta\varphi_0)
  \end{array} \right)
 \left( \begin{array}{c} x'' \\
                         y''
        \end{array} \right)
\nonumber \\ 
& &
 +R \left( \begin{array}{c} \cos (\gamma\omega t''+\Delta\varphi_0) -1 \\
                            \sin (\gamma\omega t''+\Delta\varphi_0)
           \end{array} \right) \; ,
\end{eqnarray}
\begin{equation}\label{er2''}
t=\gamma t'' +\frac{\gamma}{c^2}\omega R y'' \; .
\end{equation}
Let the labels $A$, $B$ denote the
co-ordinates of the part of the ring that lie at $S'$ and $S''$,
respectively.
Since the observer sees both parts of the ring at the same
instant, we have $t'_{A}=t'_{B}=0$. Since $x''_{B}=y''_{B}=0$, from
(\ref{er1''}) we find
\begin{equation}\label{y2}
y_{B}=R\sin (\gamma\omega t''_{B}+\Delta\varphi_0) \; ,
\end{equation}
and from (\ref{er2''})
\begin{equation}\label{t2}
t_{B}=\gamma t''_{B} \; .
\end{equation}
From $t'_{B}=0$ and (\ref{t=0})
it follows $t_{B}=\omega R y_{B}/c^2$, which, because
of (\ref{t2}), can be written as $\gamma t''_{B}=\omega R y_{B}/c^2$.
This, together with (\ref{y2}), leads to the equation that determines
$t''_{B}$:
\begin{equation}\label{eqgron11}
\gamma\omega t''_{B}=\beta^2\sin (\gamma\omega
t''_{B}+\Delta\varphi_0) \; ,
\end{equation}
where $\beta^2 \equiv \omega^2 R^2 /c^2$.
Eq. (\ref{eqgron11}) together with the fact that the rotation is 
uniform implies (\ref{eqgron1}).

To see how the inertial clock appears to the observer on the ring, 
the transformations (\ref{er1'}) and (\ref{er2'}) are sufficient.
From (\ref{er1'})
one expresses $y'$ as a function of $x$, $y$ and $t'$ and puts this
in (\ref{er2'}). The result is given by (\ref{txy}).   

Let us now discuss how the velocity of light appears to the 
observer on the ring. Let the light beam move along the gutter. 
The trajectory of
the light beam expressed in $S$-co-ordinates is given by
\begin{equation}\label{sag}
y=R \sin \omega_L t \; , \;\;\;\;\;
x=R(-1+\cos \omega_L t) \; ,
\end{equation}
where $\omega_L =\pm c/R$. The plus and minus signs refer to the
counterclockwise and clockwise propagated beams, respectively.
Using (\ref{er1'}), (\ref{er2'}), and (\ref{sag}), one can eliminate
$x,y,t$ and express $x',y'$ as functions of $t'$. The speed of
light as seen by the observer in $S'$ is
\begin{equation}\label{sag2}
v'_L =\sqrt{\left( \frac{dx'}{dt'}\right)^2 +
\left( \frac{dy'}{dt'}\right)^2 } \; .
\end{equation}
Expanding (\ref{er1'}) and (\ref{sag}) for small $t'$ and $t$, respectively,
one can easily find $y'=\pm ct'+{\cal O}(t'^2)$, $x'={\cal O}(t'^2)$,
which means that the observer sees the velocity
of light equal to $c$ when the light is at the same position
as the observer, just as expected.

\end{document}